\documentclass[pre,twocolumn,showpacs]{revtex4}
\usepackage{graphicx}
\usepackage{amssymb}
\usepackage{epstopdf}
\usepackage[usenames]{color}
\begin{document}

\title{Coalescence of low-viscosity fluids in air}
\author{Sarah C. Case}
\affiliation{The James Franck Institute and Department of Physics, University of Chicago, Chicago, Illinois 60637}

\begin{abstract}

An electrical method is used to study the early stages of coalescence of two low-viscosity drops.  A drop of aqueous NaCl solution is suspended in air above a second drop of the same solution which is grown until the drops touch.  At that point a rapidly widening bridge forms between them.  By measuring the resistance and capacitance of the system during this coalescence event, one can obtain information about the time dependence of the characteristic bridge radius and its characteristic height.  At early times, a new asymptotic regime is observed that is inconsistent with previous theoretical predictions.  The measurements at several drop radii and approach velocities are consistent with a model in which the two liquids coalesce with a slightly deformed interface.


\end {abstract}

\maketitle
 
\section{Introduction}

All around us we see fluid drops joining together:  raindrops splash into a pond and fuse with it; individual drops falling from a faucet merge together to fill a glass of water.  It is easy to forget the wonder of such a ubiquitous phenomenon.   There is a change in topology when two fluids coalesce.  As soon as they come into contact, a fluid bridge is formed between the two masses.  The initial radius of the bridge is much smaller than the macroscopic dimensions of the flow.  Interfacial tension then widens it until the two drops merge into a single entity.  A video sequence of this process is shown in Fig. 1.  Another common example of a topological transition is the inverse process to coalescence, that is, drop break up\cite{Shi_1994, Eggers_1997, Cohen_2001, Lister_1998, Chen_2002}.  There, a single mass of fluid separates into two segments joined by a thin neck.  In that case the topological transformation proceeds as a physical dimension, the neck radius, approaches zero causing the dynamics to approach a singularity.  When two drops coalesce, we expect similar singular behavior.

Such fluid transitions have often been compared to critical thermodynamic phase transitions, as this separation of length scales often leads to universal behavior\cite{Constantin_1993, Goldstein_1993, Bertozzi_1996}.  Although it is an appealing and useful framework, it was recently discovered that not all fluid-breakup singularities obey universal dynamics \cite{Doshi_2003, Keim_2006}.  In light of this, it is imperative to consider other familiar fluid transitions, such as drop coalescence, to see if they, too, behave in unexpected ways.  Moreover, drop coalescence is of practical as well as purely scientific importance.  Viscous sintering, emulsion stability and mixing in microfluidics often need to be controlled in industrial processes.   In this paper, I employ an electrical method to explore the drop coalescence transition at low viscosities at times three orders of magnitude earlier than previous optical experiments.

\begin{figure}
\centering
\includegraphics[width=90 mm]{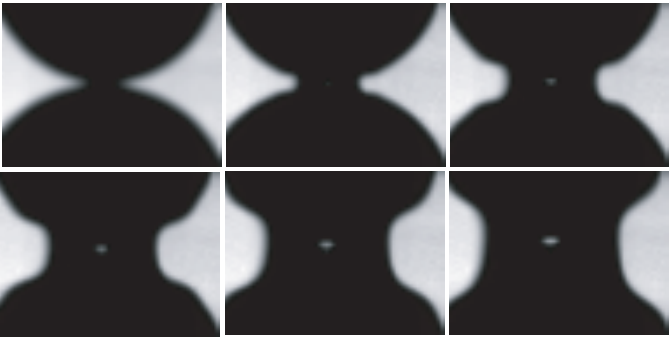}
\caption{Two drops of aqueous NaCl solution at saturation of radius $A$ = 1 mm are coalescing.  The frames are 69 $\mu$s apart.  The white spot in the bridge center is an optical artifact due to the drop lensing the light source located behind it.} 
\end{figure}

Coalescence processes occur in both the viscous regime, where the primary force opposing the widening of the bridge between the drops is due to viscous dissipation, and the inviscid regime, where the widening of the bridge is opposed primarily by inertial forces.  In the inviscid regime studied here, the radius of the fluid bridge between the two coalescing drops, $r$, is much greater than the viscous length scale of the system, $ l_{\nu} = \mu^2/\rho \gamma$,where $\mu$ is the dynamic viscosity of the fluid, $\rho$ is its density, and $\gamma$ is the surface tension.  For water coalescing in air, $l_{\nu} \approx 14$ nm, and viscous effects can be neglected for much of the coalescence.  This regime has been studied less than its high-viscosity counterpart\cite{Eggers_1999, Yao_2005} because the very rapid initial motion of the low viscosity fluid is difficult to resolve in experiments and computations.  However, theory has predictions.

A straightforward scaling argument \cite{Eggers_1999} can be used to describe coalescence in the inviscid regime.  To initiate coalescence, the drops must be brought very close together. Soon after the bridge is formed, the gap width between the two drops, $d$, will satisfy $d \ll r$.  In this case, a balance between  surface tension and inertia leads to

\begin{equation}
 (\frac{\gamma}{\rho d})^{1/2} \propto \frac{dr}{dt}
\end{equation}

It is assumed that if the two droplets are brought together sufficiently slowly, then they will maintain a spherical shape.  For hemispherical drops,  $d = r^{2}/A$, where $A$ is the drop radius, as shown in Fig. 2.  The resulting differential equation can be solved, where $t_0$ is the instant at which coalescence occurs, and c is a proportionality constant of order unity:

\begin{equation}
 r =  c(\frac{4 \gamma A}{\rho})^{1/4} (t - t_{0})^{1/2}.
\end{equation}

This scaling law has been supported by simulations studying the coalescence of low-viscosity fluid drops in the absence of an outer fluid\cite{Duchemin_2003, Lee_2006}.  However, due to the speed and the geometry of the transition, experimental studies have been unable to confirm the applicability of this scaling law to times $(t - t_0) \equiv \tau < 10$ $\mu s$.

Previous experiments have observed $r \propto (t - t_0)^{1/2}$ for  $(t - t_0) > 10$ $\mu$s, using high-speed imaging at rates up to $10^6$ frames per second \cite{Thoroddsen_2005, Menchaca_2001, Wu_2004}, as well as ultrafast x-ray phase contrast imaging\cite{Wang_2008}.  However, one aspect of the data suggests that the dynamics may not behave as we would expect.   In one experiment, a small DC voltage was placed across the drops, and it was found that the initiation of electrical contact occurred $20 - 50$ $\mu$s before coalescence could be observed visually \cite{Thoroddsen_2005}.  An electrical method introduced by Case and Nagel \cite{Case_2008}, expands the measurement range down to $\tau \sim 10$ ns, a region currently inaccessible to imaging experiments.  Results from this data closer to the instant of coalescence indicated a new asymptotic regime not predicted by the scaling argument given above.  Moreover it suggested a solution to the discrepancy found between the instant of electrical contact and the apparent initiation of coalescence. 

Electrical methods for studying drop coalescence have been used in other experiments.  The electrical method used in Case and Nagel's experiment is similar to one previously developed by Burton et al\cite{Burton_2004}, which used a small DC voltage to measure the resistance of a mercury droplet during break up.  Case and Nagel extended Burton et al's technique by using an AC voltage to measure separately both the time-dependent resistance and the capacitance of two coalescing drops of aqueous NaCl solution.  This enabled them to infer the geometry of the coalescing region as early as $10$ ns after the instant of coalescence. They found a new asymptotic regime at early times that is not consistent with the predictions of the simple scaling argument outlined above.  This behavior occurs for $\tau <  10$ $\mu$s so that it is entirely in the region that cannot be studied by direct imaging.  In addition, an AC electrical method was used by Lukyanets and Kavehpour to study the rest time of coalescing drops\cite{Lukyanets_2008}.  These results (at voltage magnitudes three orders of magnitude larger than the largest used here) suggest that deformations resulting from high electric fields in the gap between the two drops may introduce errors in measurements.  In the experiments described here, varying the voltage and frequency by several orders of magnitude does not significantly affect the results.  This is discussed further in appendix A.

In this paper, I expand upon these measurements and provide a more detailed experimental description.  I vary experimental parameters such as the drop diameter and approach velocity in order to explore further the surprising behavior seen in the initial experiments.  My experiments support the hypothesized new asymptotic regime at the earliest times measured.

\section{Experimental Description}

\subsection {Impedance Measurement}

\begin{figure}
\centering
\includegraphics[width=90 mm]{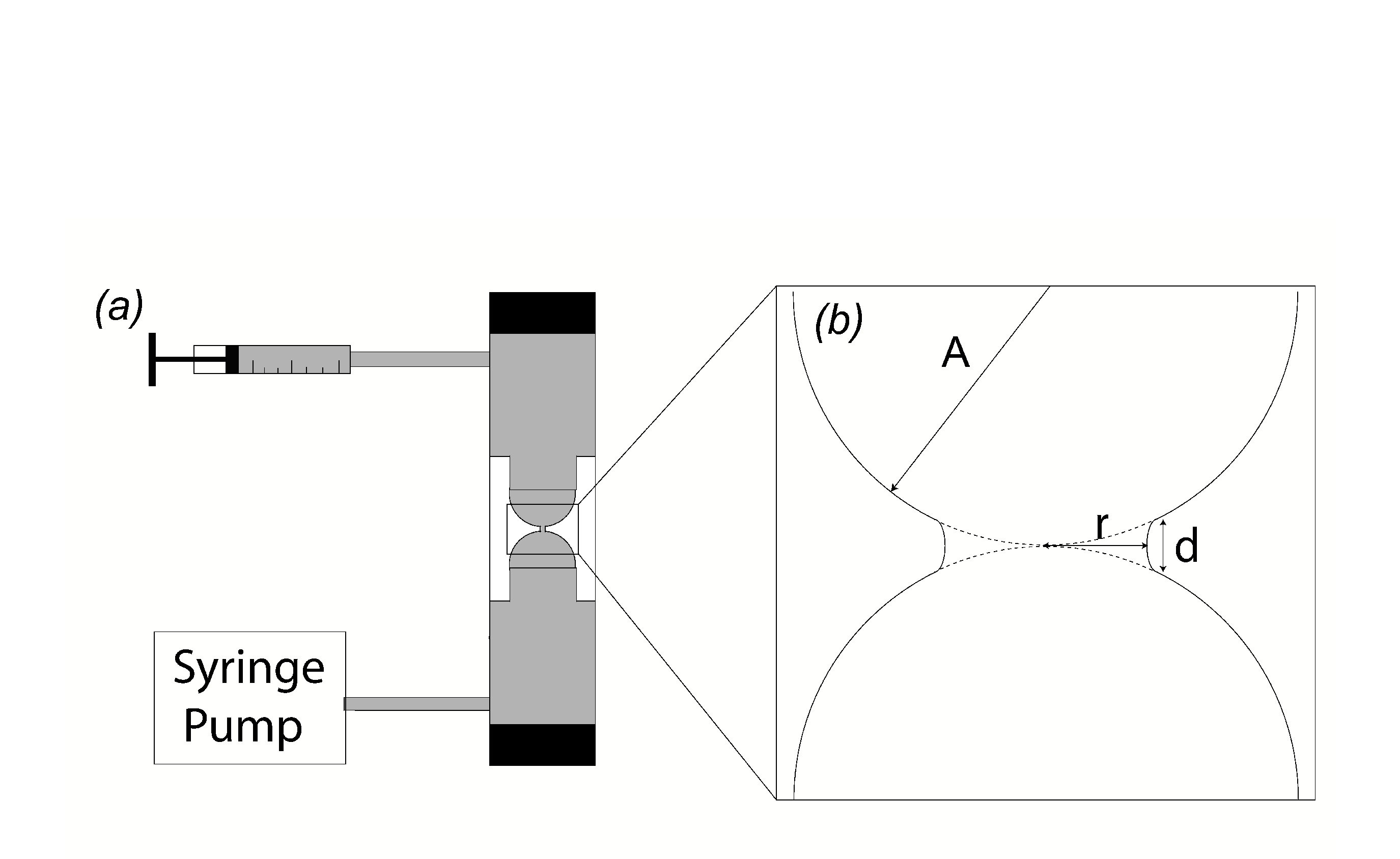}
\caption{(a) Experimental Setup. Two acrylic tubes of length 5 cm long and inner diameter 0.95 cm are secured in line with one another.  Changeable nozzles of radius $A$ are attached to each tube.  The nozzle tips are separated by $2 A$.  A drop of aqueous NaCl solution is formed on the upper nozzle using a microliter syringe, and the lower drop is then slowly grown until the two drops coalesce, using a variable speed syringe pump (Kazel R99-FM) with syringe sizes varying from 50 $\mu$l to 20 mL, and injection speeds varying from 0.21 to to 70.0 ml/hour. (b) Coalescence of two drops.  Two drops of radius $A$ meet at a single point.  A bridge of radius $r$ and height $d$ forms and expands due to the interfacial tension $\gamma$.  For hemispherical drops, $d \sim r^2/A$.} 
\end{figure} 

	In this experiment, two drops of aqueous sodium chloride solution at saturation coalesce in air at room temperature. At saturation, or $26$ $\%$ NaCl by mass, the fluid parameters of salt water are:  fluid density $\rho = 1.1972$ g/cm, kinematic viscosity $\nu = 1.662$ cSt, surface tension $\gamma = 82.55$ dyne/cm, and conductivity $\sigma = 0.225$ $(\Omega \cdot$ cm$)^{-1}$\cite{Handbook}.  
	
As shown in Fig. 2, two acrylic tubes filled with salt water were aligned vertically.  Teflon nozzles of radius $A$ were attached to each tube, facing each other.  Gold electrodes were immersed in the salt water at the end of each tube opposite the nozzle.  A known quantity of fluid was injected into each nozzle, forming two approximately hemispherical drops separated by a small distance.  An AC voltage of frequency $f$ and magnitude $|V|$ was applied across the electrodes, and the lower drop was then slowly grown at a fixed rate until the two drops coalesced.  The complex impedance of the experimental cell, $Z_{cell}$, was measured as a function of time during the coalescence.

\begin{figure}
\centering
\includegraphics[width=80 mm]{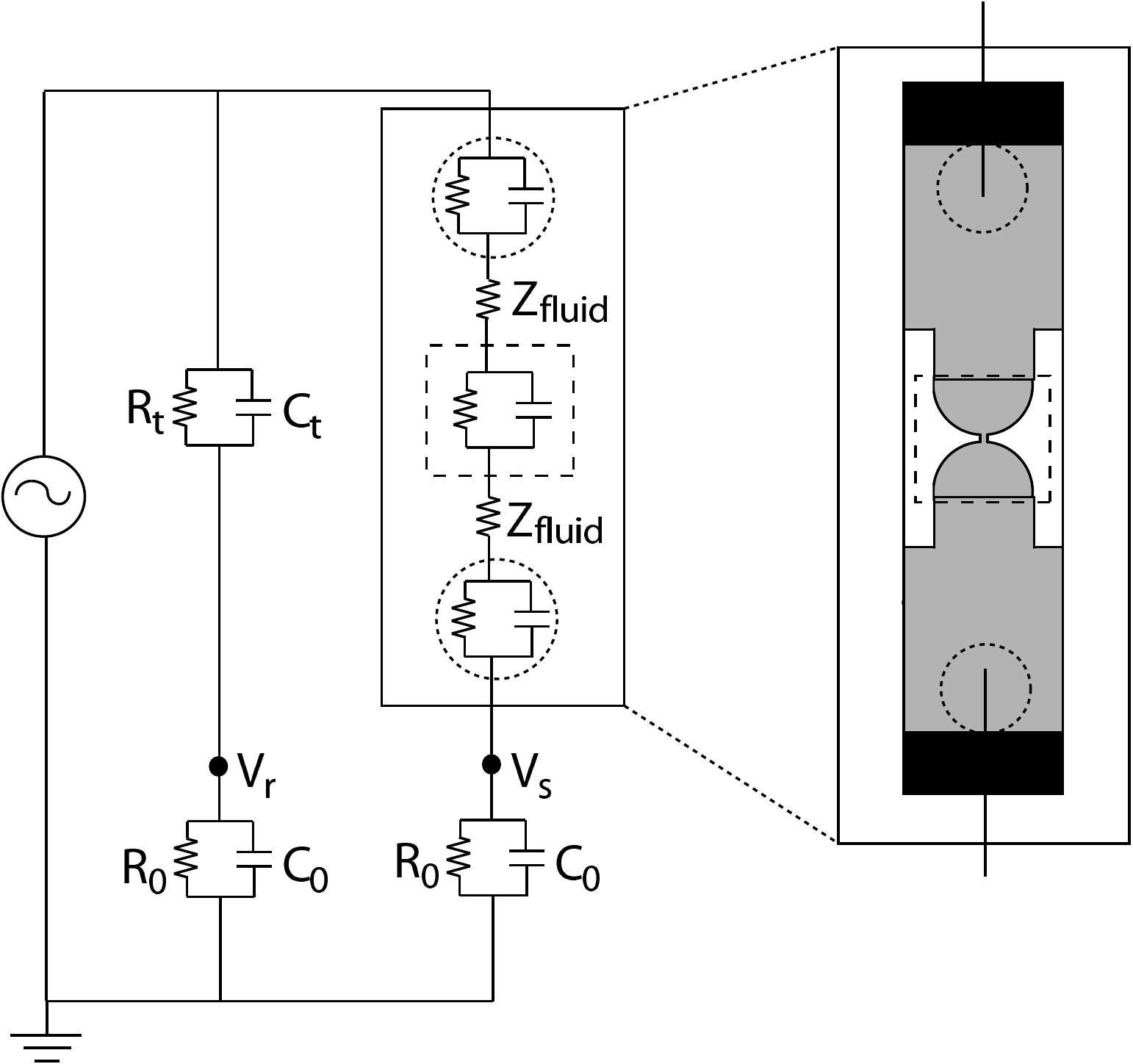}
\caption{Measurement circuit.  Gold electrodes 0.5 mm in diameter and 1 cm long are secured in the measurement cell and connected to the circuit shown.  The AC source is an HP 3325A function generator (Hewlett-Packard).  The upper left branch consists of known circuit elements ($R_t$ and $C_t$), while each lower branch is the input impedance of the oscilloscope ($R_0$) in parallel with the capacitance of the coaxial cables connecting the circuit to the oscilloscope ($C_0$).  In place of a traditional oscilloscope we use an NI PCI-5105 high-speed simultaneous sampling digitizer.  The impedance of the cell can be separated into three contributions, added in series:  $Z_{electrodes}$, $Z_{fluid}$, and $Z_{CR}$.  $Z_{electrode}$ can be modeled as a frequency-dependent capacitance due to the double layer, in parallel with an equivalent resistance due to charge transfer, as shown in the dotted circle\cite{Bockris}.  The impedance of the coalescing region, $Z_{CR}$ can be modeled as a capacitor, due to the large exposed surface of the two drops, in parallel with a resistance, as shown in the dashed square. } 
\end{figure} 
	
To evaluate $Z_{cell}$, we used the Wheatstone bridge arrangement shown in Fig. 3.  A known impedance, $Z_{t}$, was connected in series with our measuring device, a National Instruments PCI-5105 simultaneous sampling digitizer.  The effective input impedance of the PCI-5105, $Z_0$, is its stated input impedance, $R_0$, in parallel with the cable capacitance $C_0$ of the coaxial cable.  The experimental cell was also connected in series with the measuring device, and in parallel with the combination of $Z_t$ and $Z_0$.  A Hewlett-Packard HP3325A function generator was connected in series with $Z_t$.

The voltage $V_r$ was measured between $Z_t$ and $Z_0$, and $V_s$ was measured between $Z_{cell}$ and $Z_0$.  These voltages were sampled simultaneously at a maximum rate of 60 MHz.  The sampling rate was $10 f$, where $f$ is the frequency of the input sine wave from the function generator, except for data taken at $f =10$ MHz, which was sampled at the maximum rate of 60 MHz.  The voltages were read into Labview (National Instruments) and analyzed.  The analysis averaged the incoming signals over a single period to find the ratio of their amplitudes, $|V_r|/|V_s|$ as a function of time, $t$.  In addition, the analysis compared the input signals to a known sine wave and found the relative phase shift of each signal versus $t$.  These phase shifts were subtracted to find, $\Delta \phi$, the phase shift between $V_r$ and $V_s$ as a function of $t$.  

Using the complex equation:

\begin{equation}
  	\frac{|V_r|}{|V_s|} e^{- i \Delta \phi} = \frac{|Z_{cell} + Z_0|}{|Z_t + Z_0|} e^{- i \Delta \phi}
\end{equation}
allowed $Z_{cell}$ to be calculated as a function of the known circuit elements $Z_t$ and $Z_0$ and the measured values of  $|V_r|/|V_s|$ and $\Delta \phi$:

\begin {eqnarray*}
Re(Z_{cell}) & = & \frac{2|Vr|}{|Vs|} ((Re(Z_0) + Re(Z_t) )\cos \Delta \phi \\
	& & \ \ \ - (Im(Z_0) + Im(Z_t))  \sin \Delta \phi) - Re(Z_0) \\
Im(Z_{cell})  & = &  \frac{2|Vr|}{|Vs|} (((Re(Z_0) + Re(Z_t) ) \sin \Delta \phi  \\
	& & \ \ \ + (Im(Z_0) + Im(Z_t)) \cos \Delta \phi) - Im(Z_0).
\end {eqnarray*}

The circuit was calibrated by replacing $Z_{cell}$ with known circuit elements, and the measured values of $Re(Z)$ and $Im(Z)$ were shown to be consistent across the frequency range with the values of the known circuit elements.  Eq. 3 assumes that the input impedance for the PCI-5105 is identical for both input channels.  This is not necessarily the case, and the analysis allowed this to be varied in order to calibrate the cell.  Within the known error of the input impedances, however, they were identical, and this equation is accurate.

\subsection {Isolating the impedance of the coalescing region.}

\begin{figure}
\centering
\includegraphics[width=80 mm]{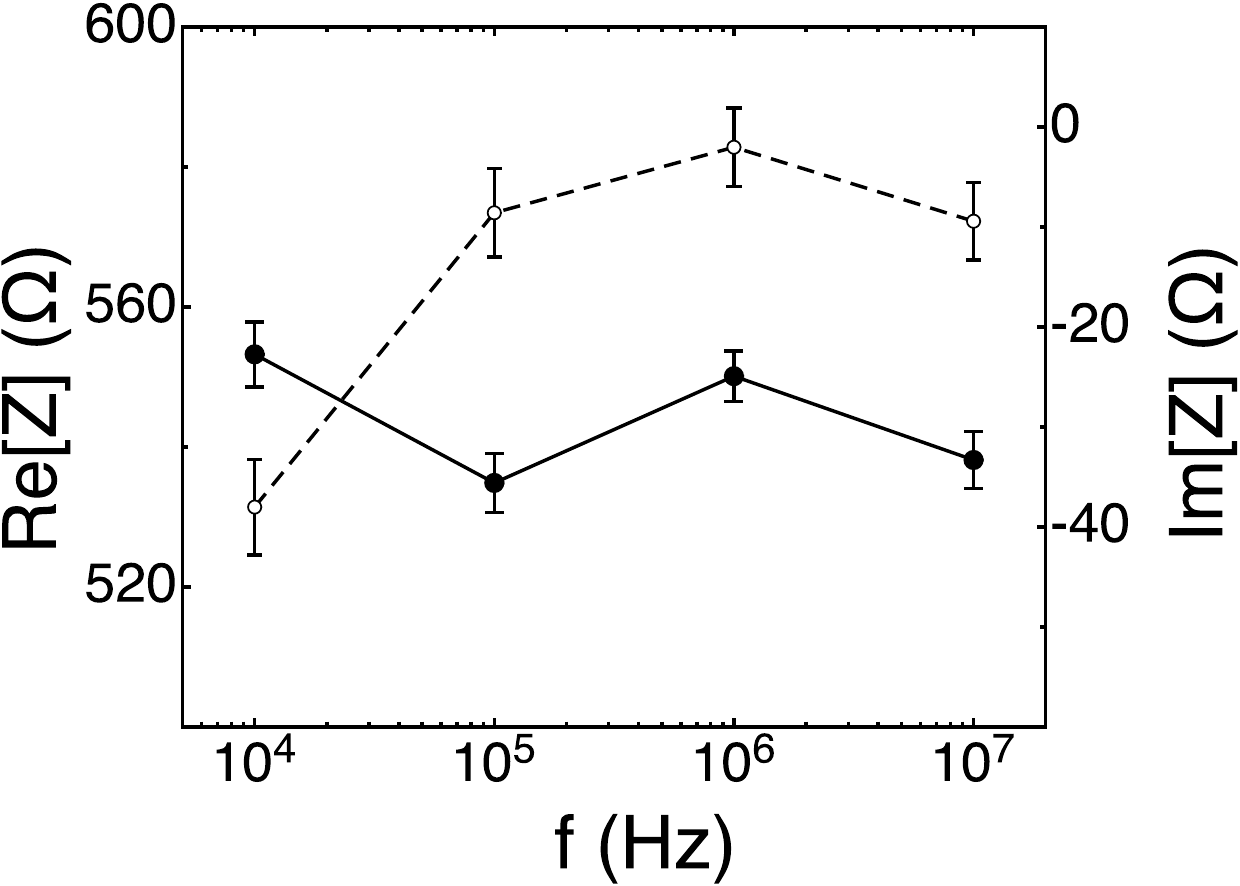}
\caption{Cell impedance. The real and imaginary parts of $Z_{closed} =  Z_{electrodes} + Z_{fluid}$ are shown as functions of frequency.  The open symbols show $Im(Z_{closed})$, and the closed symbols show $Re(Z_{closed})$.} 
\end{figure} 

The impedance of the experimental cell, $Z_{cell}$, has three distinct contributions, as shown in Fig. 3.  It can be shown that for the voltage across the cell $|V_{cell}| \lesssim 50 mV$, the interaction of the electrodes with the solution produce an effective contribution to the impedance that is equivalent to a resistor in parallel with a frequency-dependent capacitor\cite{Bockris}.  We represent this contribution as $Z_{electrodes}$.  Also, the ``Coalescing Region'' (defined as the region between the tips of the two nozzles) contributes an impedance $Z_{CR}$.  Finally, the fluid between the electrodes and the coalescing region contributes an impedance $Z_{fluid}$.  $Z_{electrodes}$ depends on the frequency $f$ while $Z_{fluid}$ is independent of $f$.  As these contributions are in series, we can write 
 \begin{equation}
 Z_{cell} = Z_{electrodes}+ Z_{fluid} + Z_{CR}.
 \end{equation}
 
If the two nozzle tips are brought into contact, $ Z_{closed} = Z_{electrodes} + Z_{fluid}$.  A representative measurement of $Z_{closed}$ as a function of frequency is shown in Fig. 4 for $A$ = 1 mm.  This measurement allows $Z_{CR}$ to be isolated from the other contributions in the cell: 
\begin{equation}
Z_{CR} = Z_{cell} - Z_{closed}.
\end{equation}

During coalescence, $Z_{CR}$ can be considered as a resistor, $R_{CR}$, representing the resistance of the two drops and the bridge between them, in parallel with a capacitance $C_{CR}$, representing the capacitance of the conducting surfaces.  Both the resistance and the capacitance change with time, and are related to the geometry of the coalescing region.  

During the 10 $\mu$s before coalescence occurs, $Z_{CR} = -i/2\pi f C_{CR}$ represents the capacitance of the drop tips as well as the other conducting surfaces of the cell.  We consider $C_{init}$, the capacitance of the two drops just before coalescence occurs, to be in parallel with the capacitance of the rest of the system, $C_{cell}$.  This approximation is supported by electrostatic simulations (see Appendix B).  Thus, in order to isolate $C_{init}$, we calculate $C_{init} = C_{CR} - C_{cell}$, where $C_{cell}$ is the measured capacitance of the cell when no drops have been formed on the nozzle tips.  In our experiments, we measure both $R_{CR}$ and $C_{CR}$ as functions of time after the bridge is formed, as well as $C_{init}$ before the bridge is formed.

\section{Results and Discussion}

\subsection{Resistance and Capacitance during Coalescence}

\begin{figure}
\centering
\includegraphics[width=80 mm]{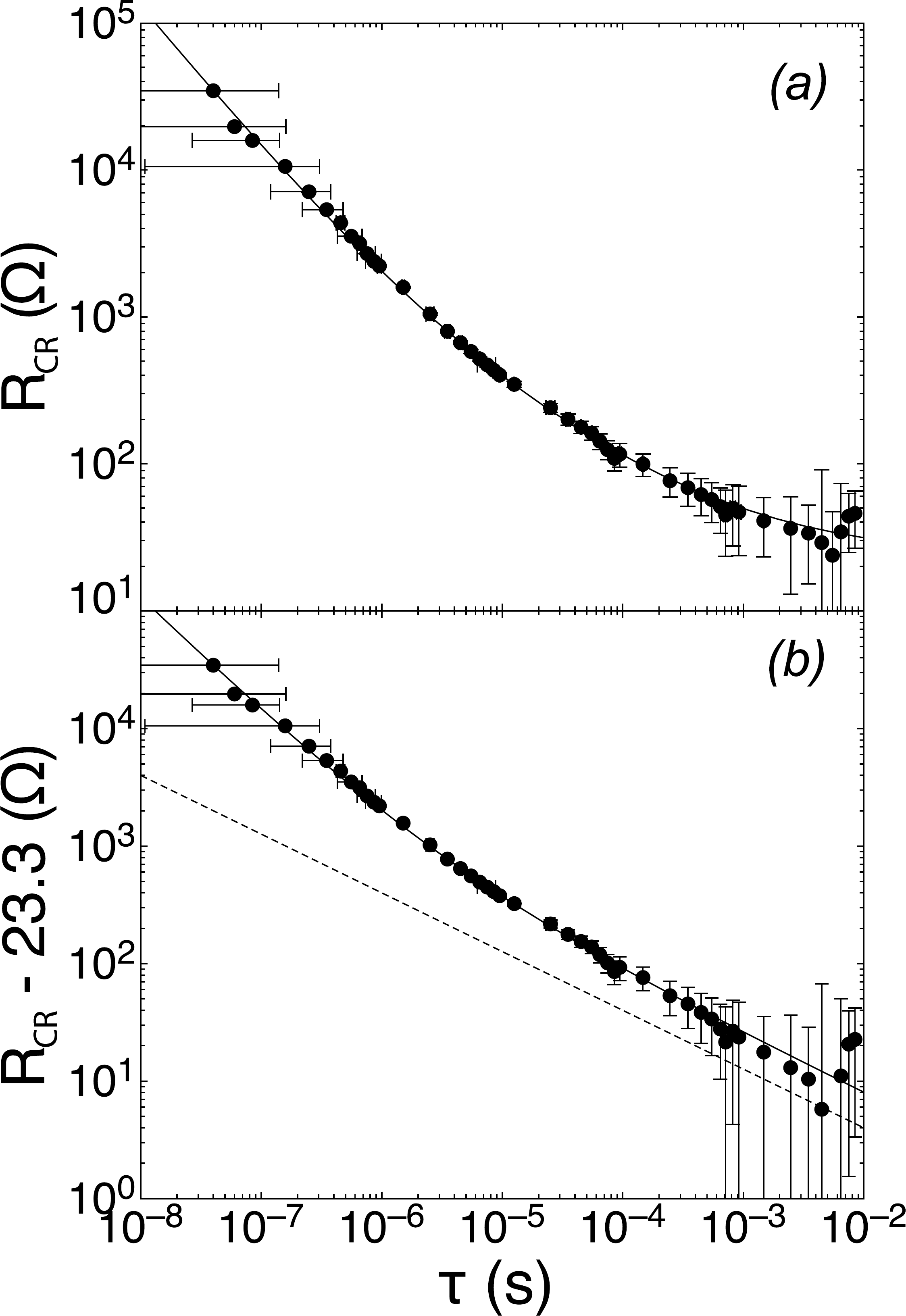}
\caption{Measured resistance during droplet coalescence.  (a)  $R_{CR}$versus $\tau = (t - t_0)$.   The solid line shows $R_{CR} = 1.2\cdot 10^{-3}\tau^{-1} + 0.8 \tau^{-1/2} + 23.3$. (b) $R_{CR} - 23.3$ $\Omega$ versus $\tau$.  The solid line shows $R_{CR} = 1.2\cdot 10^{-3} \tau^{-1} + 0.8\tau^{-1/2}$.  The dashed line shows $R_{CR} \sim \tau^{-1/2}$.  In each case, $A  = 1$ mm, and the drops approach one another at a rate of 0.0004 $A$/ms. The data is an average of 24 individual coalescence events, six obtained at each of four measurement frequencies.  The error bars reflect the spread in these measurements as well as systematic error due to inaccuracies in the measurement of $Z_{electrode}$ and due to the choice of $t_0$, the instant of coalescence. Error bars are shown both for $\tau$ and for $R_{CR}$.} 
\end{figure}
	
	The resistance of the coalescing region is shown versus $\tau \equiv t - t_0$ in Fig. 5(a) for $A = 1$ mm.  We determine the instant of coalescence ($t_0$) from the phase shift between $|V_r|$ and $|V_s|$ to within a period of the oscillation, $\pm 1/f$.  We adjust $t_0$ within this range such that the earliest data taken at a given frequency overlaps the data taken at higher $f$.  This is impossible for the highest frequency data.  
	
The data is well described by the form $R_{CR} = \alpha \tau^{-1} + \beta \tau^{-1/2} + \delta$.  The best fit to the data, shown by the solid line, gives $\alpha = (1.2 \pm 0.3) \cdot 10^{-3}$, $\beta = 0.8 \pm 0.2$, and $\delta = 23.3 \pm 15$.  Fig. 5(b) shows the same data as in Fig. 5(a), plotted as $R_{CR} - 23.3\Omega$ versus $\tau$.  The dashed line shows a power law $\tau^{-1/2}$ while the solid line shows the same fit as in Fig. 5(a).

\begin{figure}
\centering
\includegraphics[width=80 mm]{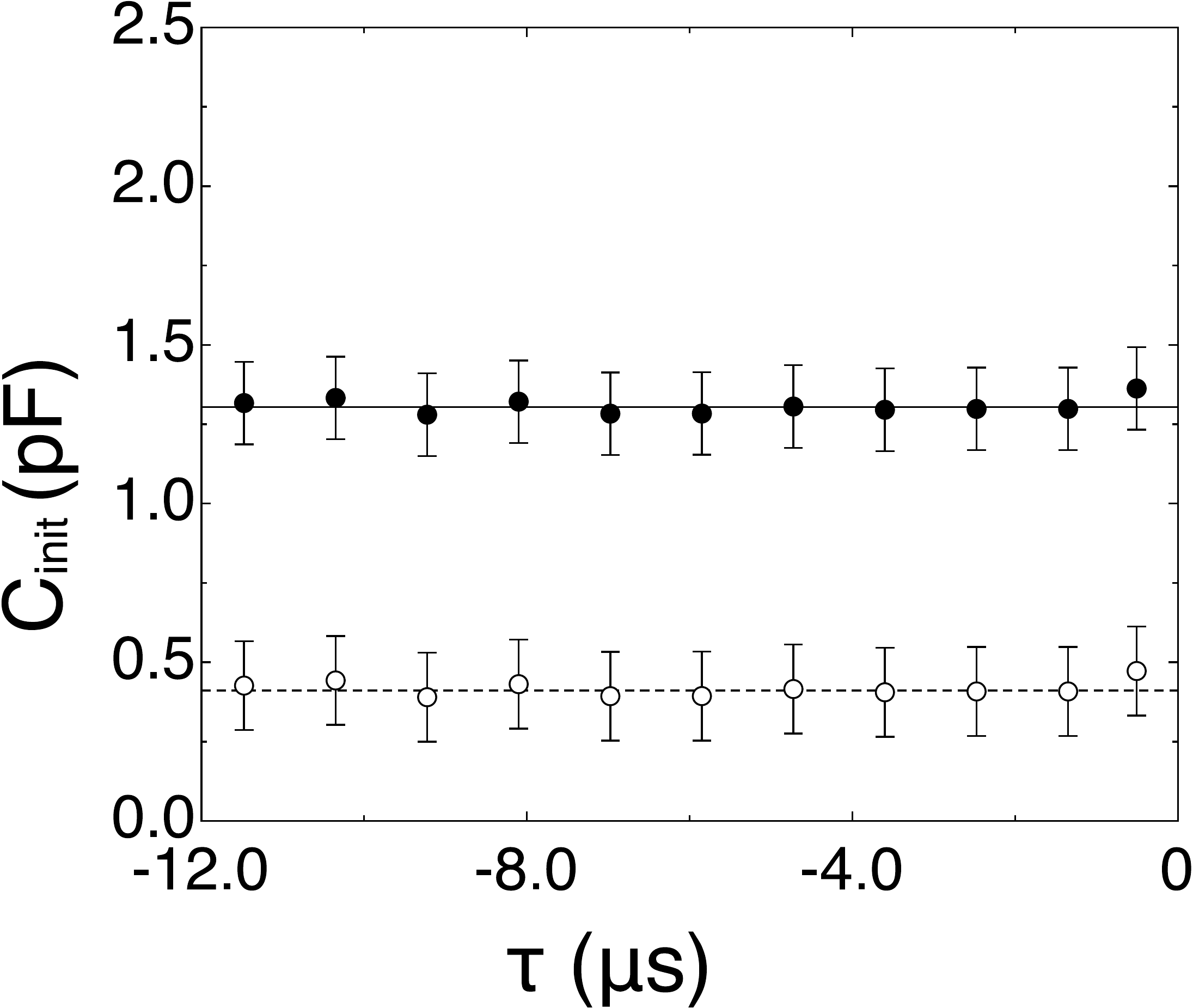}
\caption{Measured capacitance before droplet coalescence. The closed symbols show $C_{CR}$ versus $\tau$.  The solid line shows $C_{CR} = 1.3$ pF.  The open symbols show $C_{init} = C_{CR} - C_{cell}$ pF.  The dashed line shows $C_{init} = 0.41$ pF.  The data is an average of 3 individual coalescence events taken at $f = 10$ MHz.    The error bars reflect the spread in these measurements as well as systematic error. } 
\end{figure}
	
$C_{init}$ versus $\tau$ is shown in the filled symbols in Fig. 6, and is constant within error.  The solid line shows the average value of $C_{CR} = 1.30 \pm 0.14$ pF.  The open symbols show $C_{init} = C_{CR} - C_{cell}$ versus $\tau$, where $C_{cell} = 0.89 \pm 0.02$ pF.  The average value $C_{CR} = 0.41 \pm 0.14$ pF is shown by the dashed line.  All capacitance measurements are obtained at $10$ MHz, as at lower frequencies, $|V_s|$ is comparable to electrical noise.

\subsection{High-speed imaging data}

\begin{figure}
\centering
\includegraphics[width=80 mm]{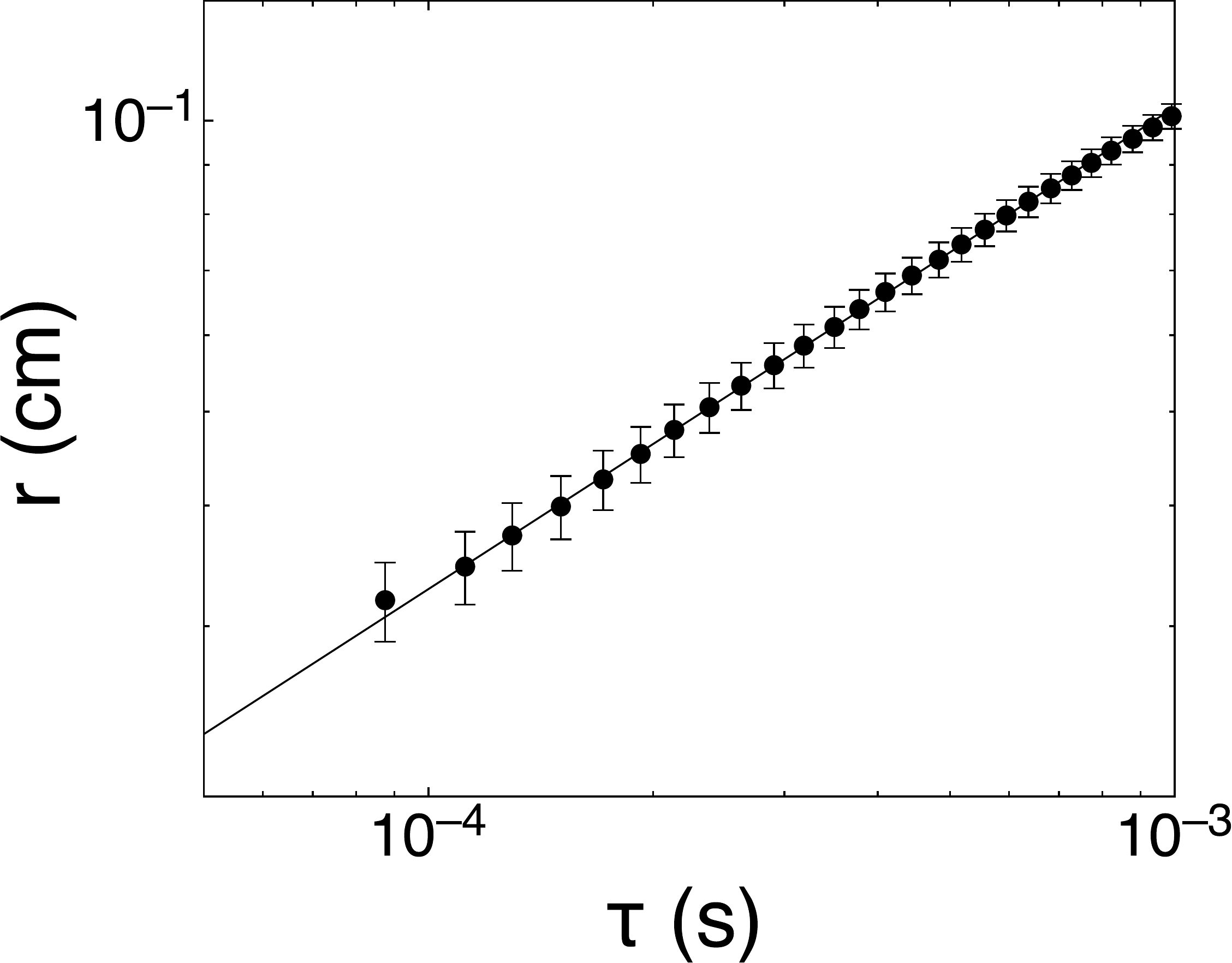}
\caption{High Speed Imaging.  $r$ versus $\tau$ is shown for a representative coalescence event.  The solid line shows $r = 3.2\tau^{0.50}$.  The frame rate is 144,000 frames per second.} 
\end{figure}

In addition to our electrical measurements, we verified previous measurements of the bridge radius during drop coalescence using a high-speed digital camera (Phantom v.7) running at 144,000 frames per second.  Images are shown in Fig. 1.   The resolution used was 26 $\mu$m/pixel, and we used simultaneous electrical measurements to determine $t_0$.  A sample measurement of $r$ versus $\tau$ for $A = 1$ mm is shown in Fig. 7.  The best fit to the data for $\tau < 1$ ms gives $r = (3.2 \pm 0.5 )\tau^{0.50 \pm 0.02}$ for $\tau > 100 \mu$s , which is consistent with previous measurements.  This exponent is consistent with the scaling argument assuming $d \propto r^2$ summarized by Eq. (2), which predicts $r = 2.3 \tau^{1/2}$.

\subsection {Comparison to predictions from scaling argument}
\begin{figure}
\centering
\includegraphics[width=80 mm]{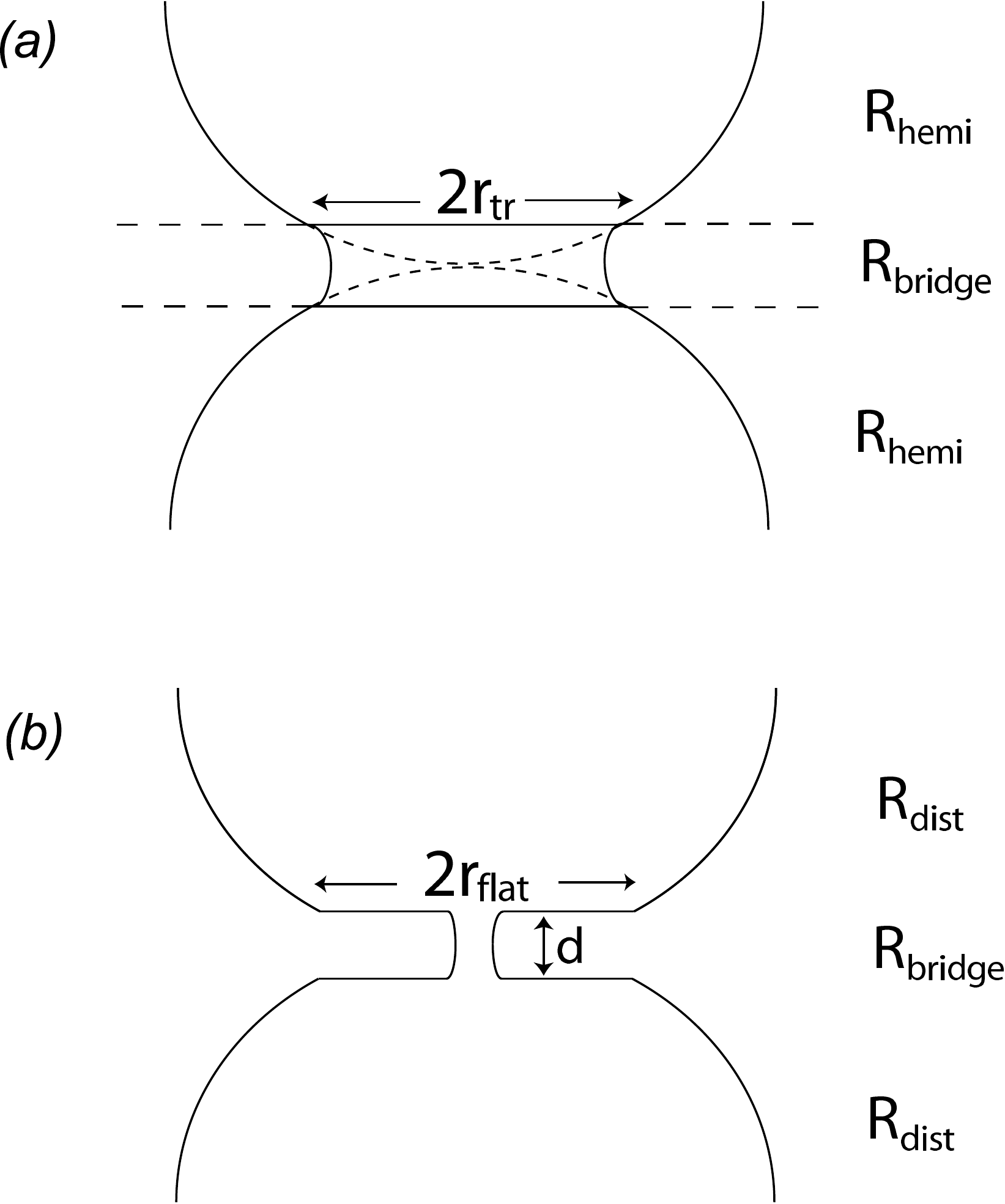}
\caption{ Two geometries for coalescence.  (a) Two hemispherical drops of radius $A$ coalesce.  We separate the resistance $R_{CR}$ into three parts. $R_{hemi}$ is the resistance of the hemispherical shapes, which are cut off as they come into contact with the bridge.  $R_{bridge}$ gives the resistance of the bridge with radius $r = r_{tr}$.  (b)  Two drops of radius $A$ coalesce with flattened tips.  The radius of the flattened region is given by $r_{flat}$.  For $r < r_{flat}$, $d =$ constant, while for $r > r_{flat}$, $d = r^2/A$.  The resistance of the flattened hemispheres is given by $R_{dist}$.}
\end{figure}

We predict $R_{CR}$ by considering the geometry of the coalescing region.  $R_{CR}$ can be separated into three pieces connected in series: $R_{CR} = R_{upper} + R_{bridge} + R_{lower}$.  $R_{upper}$ is the resistance of the upper drop, $R_{lower}$ is the resistance of the lower drop, and $R_{bridge}$ is the resistance of the bridge between the two drops.  

We assume that both drops are hemispheres of radius $A$, and so $R_{upper} = R_{lower} \equiv R_{hemi}$.  Therefore, we write $R_{CR} = 2 R_{hemi} + R_{bridge}$.  Due to the vertical alignment of the nozzles, gravity can distort the drop shapes slightly.  However, for length scales much smaller than the capillary length $l_c = \sqrt{\gamma / \rho g}$, the surface tension pressure maintaining a spherical shape is stronger than gravity.  For the aqueous NaCl solution, $l_c = 2.7$ mm, while our  largest drops have $A = 2$ mm.  

We model $R_{hemi}$ as a hemisphere truncated by a plane parallel to the flat surface of the hemisphere.  The plane intersects the hemisphere with a radius of $r_{tr}$, as seen in Fig. 8(a).  We numerically calculate the resistance of this shape using the electrostatics calculation package EStat (FieldCo), varying $r_{tr}$ over several orders of magnitude.  We find that $R_{hemi} = 1/4 r_{tr}\sigma$ , where $\sigma$ is the conductivity of the fluid.    

$R_{bridge}$ can be estimated directly.  The resistance of a roughly cylindrical object scales as the length divided by the area, yielding $R_{bridge}\sim  d/\sigma \pi r^2$.  For two hemispherical drops, the gap width  $d \sim r^2/A$,  and therefore $R_{bridge}  \sim (\sigma \pi A)^{-1}$, a constant.  Using these results and the relation between $r$ and $\tau$ from the scaling argument in Eq. (2), we find

\begin{equation}
R_{CR} =  \frac{1}{2 \sigma} (\frac{\rho}{4 c \gamma A})^{1/4} \tau^{-1/2} + \frac{1}{\sigma \pi A}.
\end {equation}	

This equation is calculated for the salt water and air system with $A = 1$ mm, $R_{CR} = 0.97 \tau^{-1/2} + 14$.  For $\tau > 10$ $\mu$s, this prediction is in qualitative agreement with our data, as can be seen in Fig. 5(b).  However, for $\tau < 10$ $\mu$s, our data shows $R_{CR}$ to be significantly larger than predicted.  Thus the early-time data suggests a new asymptotic regime not included in the scaling prediction of Eq. (2).

We can estimate $C_{init}$ in this geometry by modeling the system as two conducting hemispheres separated by a distance $z$.  When $z \ll A$, the capacitance of this arrangement of conductors is comparable to that of a sphere of radius $A/2$ suspended with its tip a distance $z$  above an infinite conducting plane, which can be solved analytically.  An approximation for $z/A \ll 1$ \cite{Boyer_1994} shows

\begin{equation}
C_{init} \approx \pi \epsilon_0 A [ \ln( \frac{A}{2z}) + 1.84].
\end{equation}

Due to the logarithmic dependence of $C_{init}$ on $z$, uncertainty in the measurement of $C_{init} = 0.41 \pm 0.14$ pF leads to enormous variation in the calculated value of $z$: from $190$ nm for $C_{init} = 0.27$ pF to $8.05 \cdot 10^{-3}$ nm for $C_{init} = 0.55$ pF.

\subsection{An Alternative Interpretation: Flattened drop tips}
To explain the discrepancy at small $\tau$ between the predictions of the model and the data, a modification to the coalescence geometry was proposed~\cite{Case_2008}.  In deriving $r \propto \tau^{1/2}$, it was assumed that $d \propto r^2$.  However, if the drop tips are slightly flattened, as in Fig. 8(b),  a different dependence is found for $r(\tau)$ at early times.  

For hemispherical drops with a flattened tip of radius $r_{flat}$, $d \propto r^2$ only when $r > r_{flat}$.  For $r < r_{flat}$, $d$ is constant, and the problem is equivalent to that of a hole opening in a thin film due to interfacial tension.  As long as we remain in the inviscid regime, Eq. (1) still applies, and solving for $d$ constant, we find 

\begin{equation}
        r = c' ( \frac{\gamma}{\rho d})^{1/2} \tau.
\end{equation}
In this geometry, $R_{CR}$ is calculated by replacing $R_{hemi}$ for the undistorted case with $R_{dist}$, the resistance of the distorted hemisphere in Fig. 8(b) that has a flattened tip of radius $r_{flat}$.  The flattened hemisphere has a small opening in its base of radius $r$ which corresponds to the bridge.  A numerical solution shows $R_{dist} = 1/4 r \sigma$.  We estimate $R_{bridge}=  d/\sigma \pi r^2$.  Combining these contributions with the time dependence seen in Eq. (8) yields
\begin{equation}
R_{CR} = \frac{1}{2 \sigma} (\frac{\rho}{\gamma})^{1/2} \frac{d^{1/2}}{\tau} + \frac{\rho}{\sigma \pi \gamma} \frac{d^2}{ \tau^{2}}.
\end{equation}
Thus, when $r < r_{flat}$, $R_{CR}$ is independent of $A$.  

The predictions of this model can be compared to the data shown in Fig. 5(a).  A transition time $t_t$ from the $\tau^{-1/2}$ behavior to the $\tau^{-1}$ behavior is determined to be $0.87$ $\mu$s $\le t_t \le 6.5$ $\mu$s with the best fit being  $t_t = 2.4$ $\mu$s.  For $t \ll t_t$, $R_{CR} \approx  (1.2\cdot 10^{-3} \pm 3 \cdot 10^{-4}) \tau^{-1}$.  From the argument above, this prefactor is $(\rho d/\gamma)^{1/2}/2 \sigma$.  Comparing the prediction to the data yields $d = 200 \pm 100$ nm. 

The contribution from the $\tau^{-2}$ term is negligible.  $R_{CR}$ crosses over from $\tau^{-1}$ behavior to $\tau^{-2}$ behavior at a time $t_c = (0.077$ s/cm$^{-3/2}) d^{3/2}$.  When $d = 200 \pm 100$ nm, $t_c = 7 \pm 5$ ns.  This is beyond the measurement window; the experiments would only resolve $R_{CR} \propto d^{1/2}/\tau$ at the earliest times measured.  

In this model, $C_{init}$ is dominated by the flattened region.  We approximate $C_{init}$ as a parallel plate capacitor of area $\pi r_{flat}^2$ and separation $d$.  Using Equation (2) and the crossover time $t_t$ from $\tau^{-1/2}$ to $\tau^{-1}$ behavior gives $22$ $\mu$m $\le r_{flat} \le 59$ $\mu$m with the best fit value (from the best fit for $t_t$ above) $ r_{flat} \approx  36$ $\mu$m.  This approximation then yields a capacitance $0.04$ pF $\le C \le 0.97$ pF with the best fit $C = 0.18$ pF.  Using electrostatic simulations in combination with Eq. (7), I find that the hemispherical region contributes approximately $0.07$ pF to the capacitance.  Assuming these parallel contributions can be added, I find that $0.11$ pF $\le C_{CR} \le 1.04$ pF, which is consistent with the measurement $C_{init} = 0.41$ pF.

\subsection{Varying the drop radius $A$}

\begin{figure}
\centering
\includegraphics[width=80 mm]{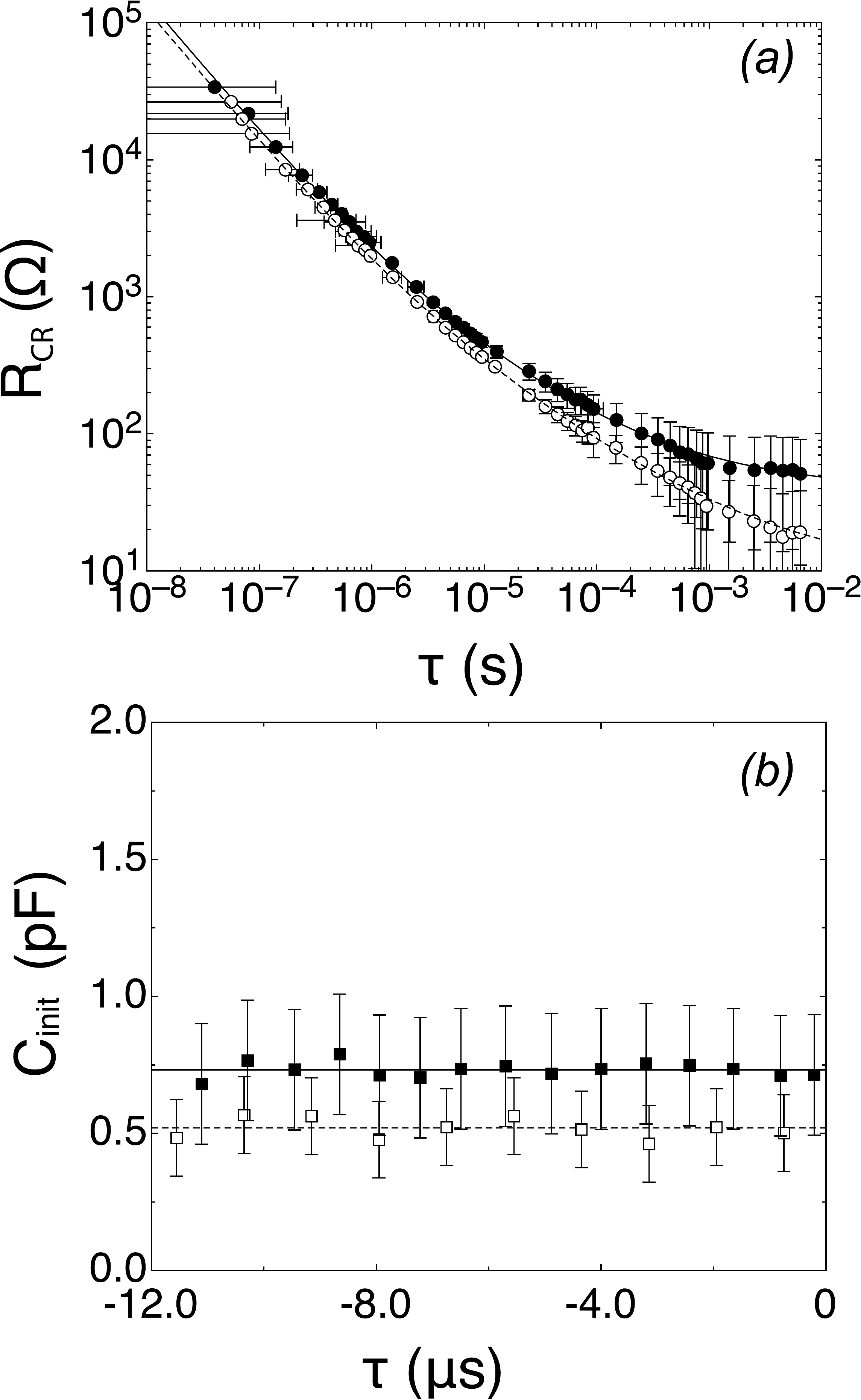}
\caption{Varying drop radius.  The open symbols show $A$ = 2 mm.  The closed symbols show $A$ = 0.75 mm. (a) $R_{CR}$ versus $\tau$. Data shown is the average of 12 individual coalescence events for each $A$.  Three coalescence events were measured with each of four different $f$, from $f = 10$ kHz to $10$ MHz.  The solid line shows $R_{CR} = 1.4 \cdot 10^{-3} \tau^{-1} + 0.9 \tau^{-0.50} + 40$. The dashed line shows $R_{CR} = 1.2 \cdot 10^{-3} \tau^{-1} + 0.7\tau^{-0.50} + 10$.   (b) $C_{init}$ versus $\tau$.  Data shown is the average of 3 coalescence events taken at $f = 10$ MHz.  The dashed line shows $C_{init} = 0.52$ pF.  The solid line shows $C_{init} = 0.73$ pF.} 
\end{figure} 

The model with the flattened tips has two regimes.  Eq. 9 should hold for $t \ll t_c$, where $t_c$ represents the time at which $R_{CR}$ crosses over from $\tau^{-1}$ to $\tau^{-2}$ behavior.  At longer times,  $t \gg t_c$, $R_{CR}$, Eq. 6 should apply.   Varying the drop radius $A$ should only affect data for $t \gg t_c$.  

I have measured $R_{CR}$ for $A =2$ mm and $A = 0.75$ mm in addition to the $A =1$ mm measurements already shown.  For a drop with a flat tip, we expect that for $t \gg t_t$, $R_{CR} \approx 0.8 \tau^{-1/2} + 7$ for $A = 2$ mm, and $R_{CR} \approx 1.0 \tau^{-1/2} + 19$ for $A = 0.75$ mm.   At early times, we expect no change outside of error from the $A = 1$ mm data for both drops. 

The measured $R_{CR}$ versus $\tau$ for $A =2$ mm and $A = 0.75$ mm is shown in Fig. 9(a).  For $A =2$ mm (open symbols), the best fit to the data is the dashed line: $R_{CR} = (1.2 \pm 0.3) \cdot 10^{-3}\tau^{-1} + (0.7 \pm 0.2) \tau^{-0.50} + (10 \pm 10)$.  For $A$ = 0.75 mm (closed symbols), the best fit is the solid line: $R_{CR} = (1.4 \pm 0.3) \cdot 10^{-3} \tau^{-1} + (0.9 \pm 0.2) \tau^{-0.50} + (40 \pm 15)$.   Our measurements thus give prefactors that are qualitatively consistent with those predicted by the flattened tip model.

$C_{init}$ versus $\tau$ is shown for $A = 2$ mm and $A = 0.75$ mm in Fig. 9(b).  $C_{cell}$ for $A = 2$ mm is measured to be $0.92$ pF, and $C_{cell}$ for $A = 0.75$ mm is $0.81$ pF.  The measured capacitance for $A = 0.75$ mm (shown in closed symbols) is larger than that seen for $A = 1$ mm, and the average value before coalescence is $C_{init} - 0.84 = 0.73 \pm 0.2$ pF (shown by the solid line.)  For $A = 2$ mm, we observe $C_{init} - 0.92 = 0.52 \pm 0.14$ pF, which is within error of the value observed for $A = 1$ mm.

In summary, when $A$ is varied, we observe behavior which is consistent with the coalescence of two slightly flattened drops.  For $\tau < 1$ $\mu$s, no difference is observed in $R_{CR}$ when $A$ is increased by a factor of $2.7$.  For $\tau \gg 1$ $\mu$s, the observations are consistent with the prediction that $R_{CR}$ should increase for smaller drops.   In the capacitance measurements, we observe increases over the $1$ mm measurement for both $A = 0.75$ mm and $A = 2$ mm.  An increase in deformability could account for the increase observed for $A = 2$ mm.  For $A = 0.75$ mm, we are unable to reach the low approach velocities used for larger drops, which may be responsible for the large capacitance observed.  

\subsection{Varying the Drop Velocity}

\begin{figure}
\centering
\includegraphics[width=80 mm]{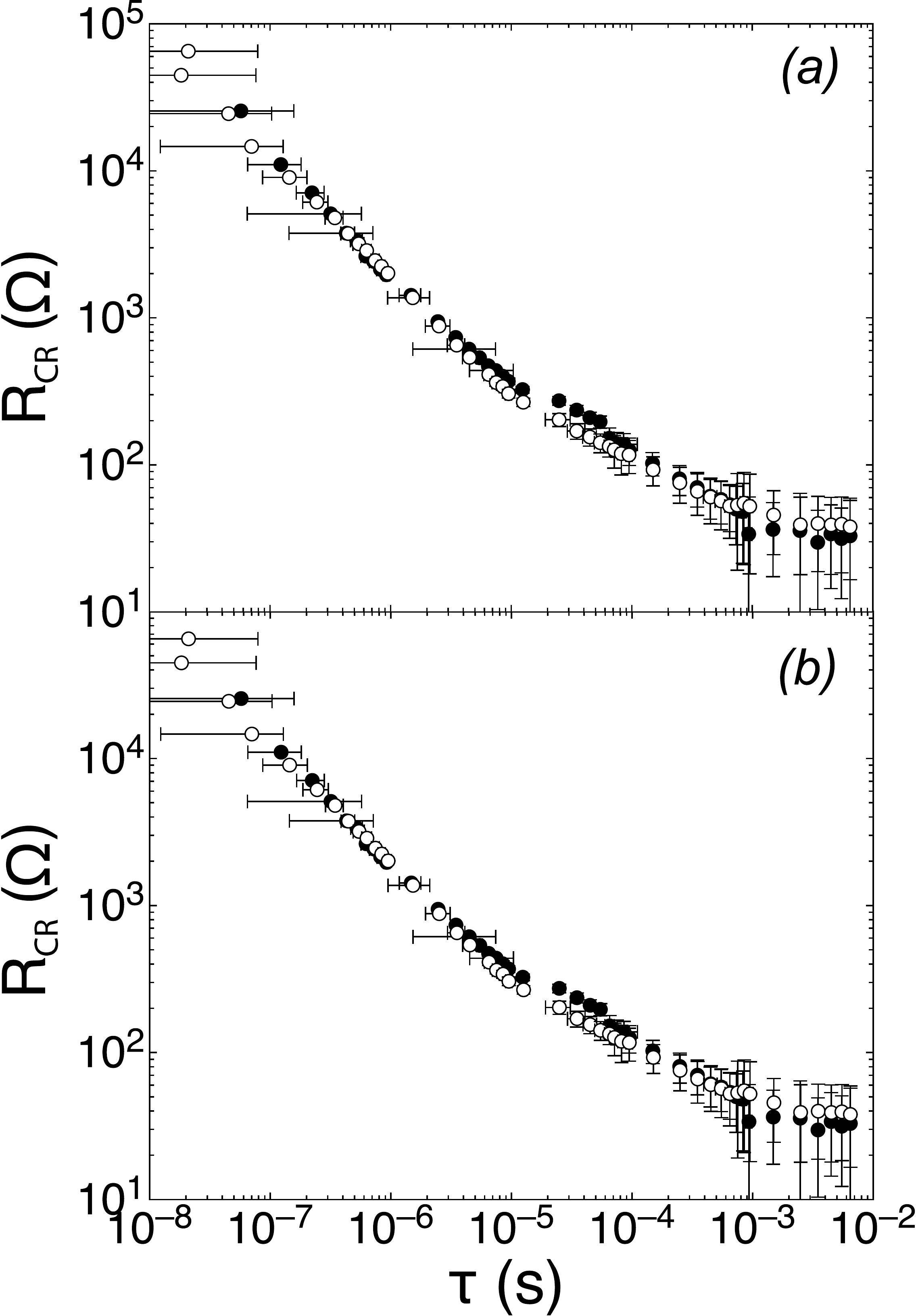}
\caption{Varying approach velocity.  (a) $R_{CR}$ versus $\tau$ for $A = 1$ mm. The closed symbols show $v = 0.0004$ $A$/ms.  The open symbols show $v = 0.0020$ $A$/ms.  Data shown is the average of 12 individual coalescence events for each $v$.  Three coalescence events occurred at each of four different $f$, from $f = 10$ kHz to $10$ MHz.  (b) $R_{CR}$ versus $\tau$ for $A = 2$ mm. The closed symbols show $v = 0.0001$ $A$/ms.  The open symbols show $v = 0.0017$ $A$/ms.} 
\end{figure} 

The drops are brought together at a non-negligible approach velocity $v$.  As $v$ is increased, air effects will become more marked, particularly for larger drops, and may change the drop shape.  We isolate such effects by varying the approach velocity.  To account for the effects of different $A$, our units of velocity are $A$/ms.

We show $R_{CR}$ versus $\tau$ in Fig. 10(a), where the approach velocity is varied by a factor of 5 for $A = 1$ mm.  The closed symbols represent $v = 0.0004$ $A$/ms, while the open symbols represent $v = 0.002$ $A$/ms.  Varying the velocity by this amount does not appreciably change the average data. 

Fig. 10(b) shows $R_{CR}$ versus $\tau$ for $A = 2$ mm, where the approach velocity is varied by a factor of 17.  In this case, the closed symbols represent $v = 0.0001$ $A$/ms, and the open symbols represent $v = 0.0017$ $A$s/ms.  The data for $v = 0.0004$ $A$/ms, shown previously, is within error of the $v = 0.0001$ $A$/ms data.

\begin{figure}
\centering
\includegraphics[width=80 mm]{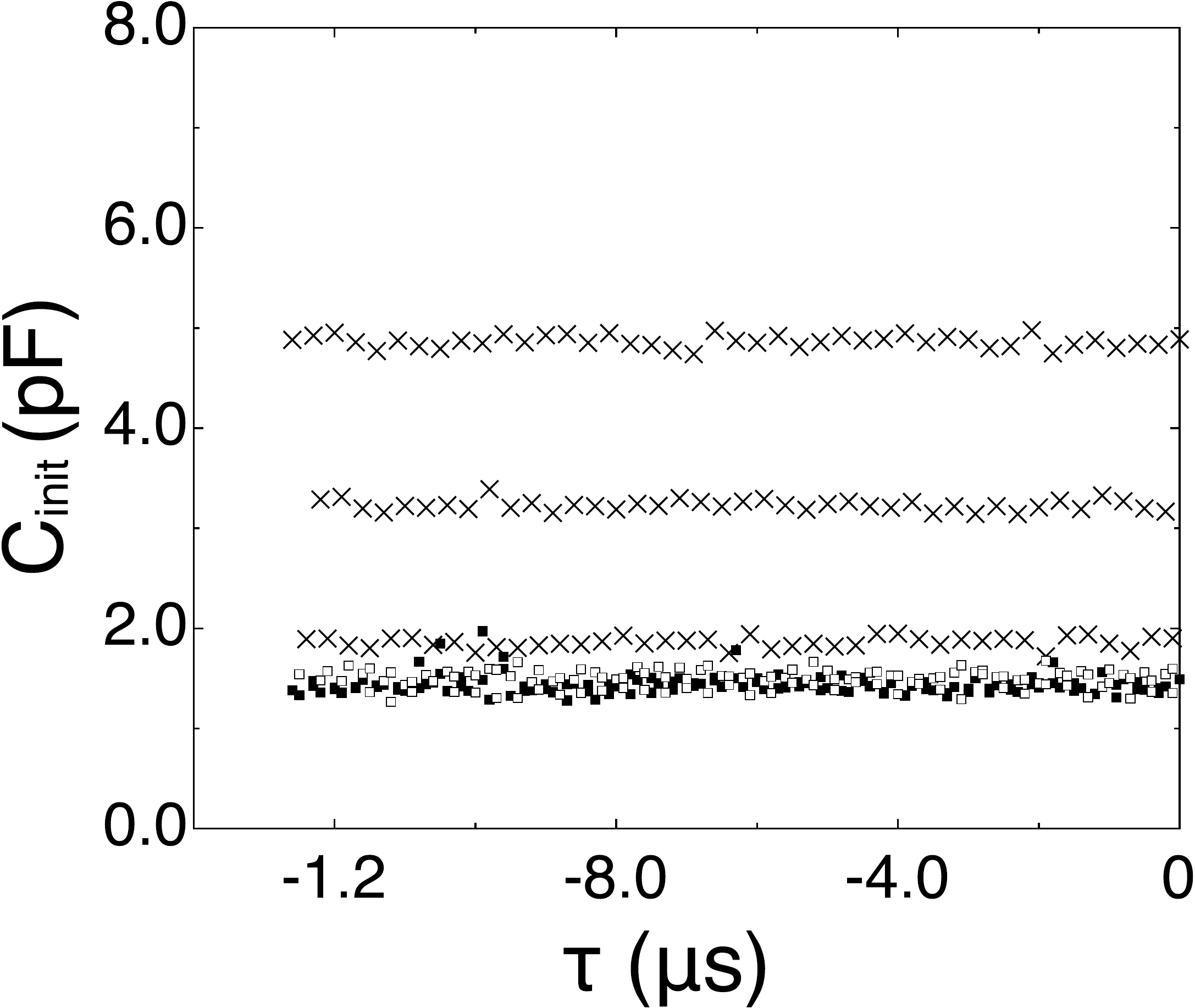}
\caption{Varying approach velocity.  $C_{init}$ versus $\tau$.  Closed symbols show data from three separate coalescence events for $v = 0.0004$ $A$/ms.  Open symbols show data from three separate coalescence events for $v = 0.0001$ $A$/ms.  The crosses show data from three separate coalescence events for $v = 0.0017$ $A$/ms.  All data shown was measured at $f = 10$ MHz.} 
\end{figure} 

Fig. 11 shows $C_{init}$ versus $\tau$ for $v = 0.0001$ $A$/ms, $v = 0.0004$ $A$/ms, and $v = 0.0017$ $A$/ms, where $A = 2$ mm.  Three independent coalescence events are shown for each velocity.  For  $v = 0.0001$ $A$/ms and $v = 0.0004$ $A$/ms, the capacitance is very reproducible, and $C_{init} =  0.52 \pm 0.14$ pF.  However, at  $v = 0.0017$ $A$/ms, the capacitance is significantly larger and also less reproducible between different events.  

We do not see a significant change in the data when we increase the approach velocity up to $v = 0.002$ $A$/ms for $A = 1$ mm and up to $v = 0.0004$ $A$/ms for $A = 2$ mm.  The behavior we see for drops $A = 2$ mm, $v = 0.0017$ $A$/ms is consistent with increased deformation of the drops.  A highly deformed drop would coalesce at the same $d$ as a less deformed drop, but $t_c$ would occur later, and the effective $A$ could be expected to be larger.  This leads to a lower $R_{CR}$ than a hemispherical drop for $r > r_{flat}$, which we see for $A = 2$ mm and $v = 0.0017$ $A$/ms when compared to $v = 0.0001$ $A$/ms.  A large deformation would also increase $C_{init}$ significantly, which we also observe in the highest-velocity data for $A = 2$ mm.  The lack of reproducibility of $C_{init}$ for $v = 0.0017$ $A$/ms may indicate that the radius of the flattening is not consistent between different events at large velocity.

\section{Conclusions}

In conclusion, an electrical method has been used to study the coalescence of two salt water drops.  This method allows us to observe an unexpected asymptotic regime which becomes visible at $\tau < 10$ $\mu$s.  Our data is consistent with the coalescence of two slightly flattened hemispherical drops.  This is contrary to previous expectations, in which the drops were expected to maintain shapes described by quadratic minima.  In addition, when $A$ is varied by nearly a factor of three, we continue to observe behavior which is consistent with the coalescence of two slightly flattened drops.  Within error, we do not see a significant change in the data when we increase the approach velocity.  

A previous theoretical description\cite{Duchemin_2003} has suggested that coalescence may occur as capillary waves cause repeated connections of the gap between the two drops.  Each connection would entrain a toroidal bubble of the outer fluid.  We see no evidence of this behavior in these experiments, which would appear as discrete jumps in $R_{CR}$ during coalescence, as the neck width widens at each connection.  It is possible that this behavior occurs on a timescale that is faster than the experiments described here are able to resolve.

Previous experiments using high-speed imaging have been unable to resolve this early-time regime.  Additionally, they found that electrical contact occurred $20$ to $80$ $\mu$s before the first motions of coalescence were observed visually\cite{Thoroddsen_2005, Menchaca_2001}.  If the drops are coalescing in a flat region with $d \sim 100$ nm, it would be impossible to observe this stage of coalescence visually.  It would only be possible to observe coalescence when $d$ increased, entering into the hemispherical regime.  We find that this occurs between $1$ and $10$ $\mu$s after the initiation of coalescence, consistent with these observations.  These previous experiments also postulated that coalescence occurred over a finite region of radius $ \sim 100$ $\mu$m.  In this case, $R_{CR}$ would increase suddenly at the instant of coalescence, contrary to our observations.  
		
There are several possible reasons for the existence of a flattened region, and I suggest two here.  One possibility is that the flattening is an air effect.  It has recently been shown that for drops splashing on dry surfaces, air plays a role in the dynamics of the impact\cite{Xu_2005}.  Two drops approaching one another at finite velocity might trap a layer of air between them, which could have unexpected consequences.  A second possibility is the presence of surfactant.  Although precautions were taken to avoid contamination\footnote{Fresh fluid was used before each data set was taken.  Also, the experiment was cleaned before each day of data taking.}, surfactants might still be present in small quantities.  It has been seen that even a very small amount of surfactant can prevent a drop from coalescing with a flat fluid surface\cite{Amarouchene_2001}.  The repulsion due to this could explain the observed flattening.
	
Understanding the flattening of two fluid drops as they approach each other could not only affect the many industrial applications that rely on droplet coalescence, but also could illuminate other important physical questions.  The origin of the thin film rupture which triggers coalescence is an active field of research\cite{Oron_1997}.  The observable flattening of the drops could contribute to an understanding of this rupture.  Also, as we begin to study the topological changes that occur in microfluidics, behavior at the smallest scales and the earliest times is essential.
		
The earliest stages of a topological transition in a fluid are when the analogy to a critical thermodynamic phase transition ought to be most accurate.  As in drop break up, near the coalescence transition, the small-scale flows decouple from the large-scale flows.  However, we observe that the bridge radius between the two drops scales differently depending on the overall drop shape.  The geometry of the system is crucial, a situation that does not have an analog in thermodynamic phase transitions.  By studying these fluid shape transitions, we widen our understanding of the many unexpected ways in which nature produces these remarkable transformations.

	I am grateful to S. R. Nagel, L. N. Zou, X. Cheng, J. L. Wyman, N. Keim, E. Corwin, and J. Royer for helpful discussions and feedback.  This research was supported by NSF MRSEC DMR-0213745 and NSF DMR-0652269.

\appendix
\section{Checks on Method}

Our experimental method, although similar to previous methods, has many novel elements.  We present here several checks that we have performed to validate our data.  This includes isolating electrical effects as well as using our method to study a better-understood topological change, that of drop snap-off of water in air.

\subsection{Varying electrical parameters}

\begin{figure}
\centering
\includegraphics[width=80 mm]{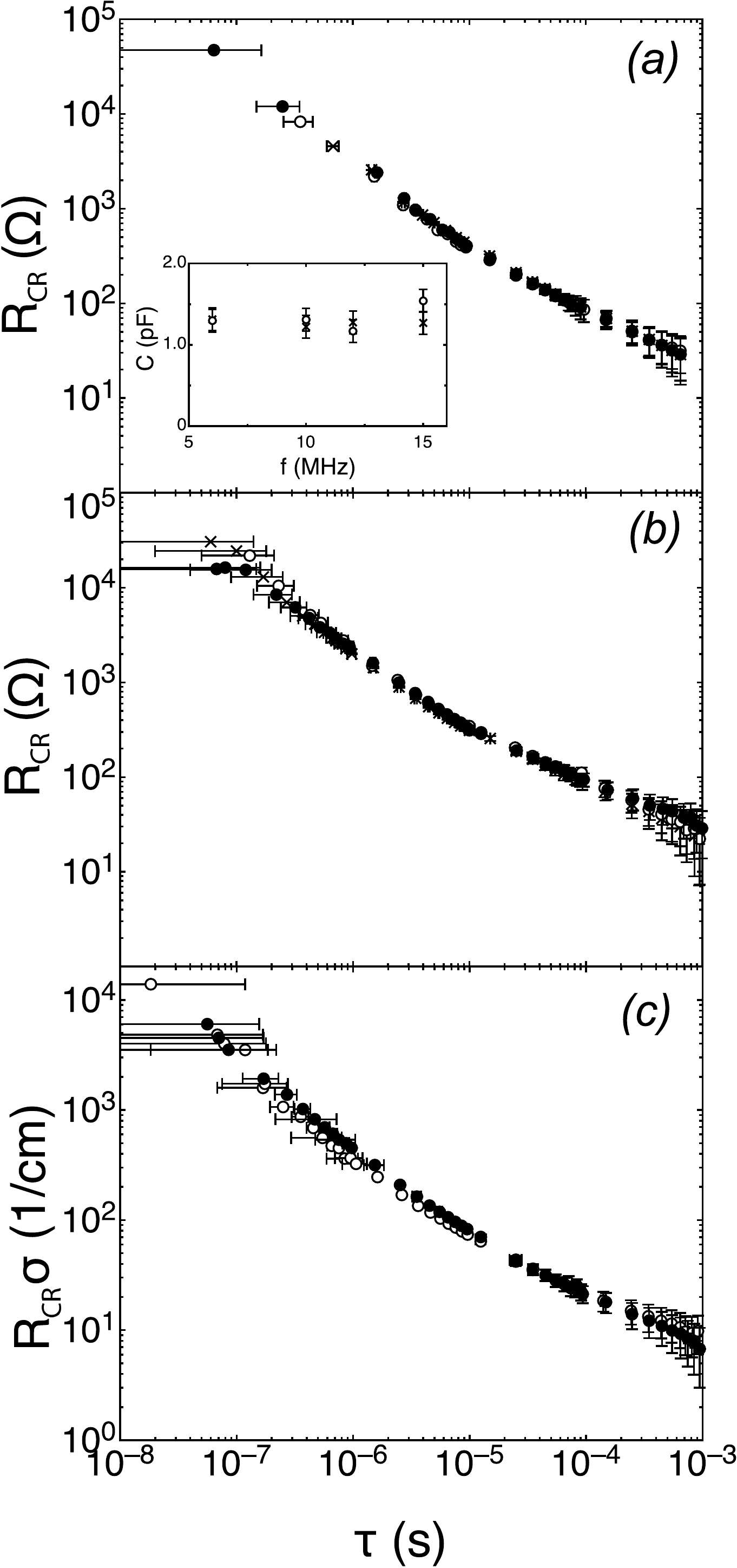}
\caption{Varying electrical parameters in resistance measurements.  Each set of data shown is the average of 12 individual coalescence events.  (a) Varying $|V_{max}|$.  $R_{CR}$ versus $\tau$ is shown.  Closed symbols show $|V_{max}| = 25$ mV.  Open symbols show $|V_{max}| = 50$ mV.  Crosses show $|V_{max}| = 250$ mV.  Inset shows $C_{init}$ versus $\tau$.  Open symbols show $|V_max| = 50$ mV.  Crosses show $|V_{max}| = 250$ mV. (b)  Varying DC component.  $R_{CR}$ versus $\tau$ is shown.  Closed symbols show no added DC.  Open symbols show an added DC component of $140$ mV.  Crosses show an added DC component of $35$ mV.  (c) Varying ionic concentration.  $R_{CR}/\rho_r$ versus $\tau$ is shown.  Closed symbols show a $26 \%$ by mass solution (saturation), $\sigma = 0.225$ $1/\Omega \cdot$ cm.  Open symbols show a $10 \%$ by mass solution, $\sigma = 0.126$ $1/\Omega \cdot$ cm.} 
\end{figure} 

\subsubsection{Varying Voltage Magnitude}
We begin by verifying that our measurements do not change if the magnitude of the AC voltage across the cell is changed.  For $|V| \gtrsim 50$ mV, we cannot approximate the electrodes as a resistor in parallel with a capacitor, as the charge transfer reaction no longer gives us $I \sim V$.  In addition, large $|V|$ could deform the shape of the drops by increasing the attraction or repulsion between the surfaces.    

During coalescence, as $Z_{cell}$ changes, the voltage across the cell, $V_{cell}$, also changes in magnitude.  We can determine a maximum $|V_{cell}|$, $|V_{max}|$, by assuming that the full voltage drop supplied by the function generator occurs across the experimental cell.  We vary $|V_{max}|$ from 12.5 mV to 500 mV, and find no significant difference in our data, as shown in Fig. 12(a) for three sample $|V_{max}|$.  The noise becomes comparable to our signal at 12.5 mV, and all data used in the analysis is taken at $|V_{max}| = 50$ mV.

Additionally, as shown in the inset to Fig. 12(a), we examine the effect of varying $|V_{max}|$ from $50$ mV to $500$ mV on the measurement of $C_{init}$ using measurement frequencies from $f = 6$ MHz to $15$ MHz.  No difference is seen outside of error.

\subsubsection{Varying DC component of input signal}

In addition to the AC signal applied across the experimental cell, a small amount of DC signal is observed.  Any DC signal applied across the experimental cell before coalescence will polarize the cell, effectively charging it up like a capacitor.  At the instant of coalescence, this capacitor discharges, and a DC spike is observed.  This spike has a typical maximum size $ \sim 1-10 \mu$V, which is less than $1 \%$ of the typical output signal.  We average the signal and remove this DC contribution before analysis.  We check the validity of this by explicitly adding up to 140 mV of DC to our signal.  This did not alter our measurement within error, as can be seen in Fig. 12(b).

\subsubsection{Varying ionic concentration}

All data presented in the main body of the paper was taken with NaCl in water at saturation, or $26 \%$ by mass.  We took the same set of measurements with solution of NaCl $15 \%$ by mass and $10 \%$ by mass.  The conductivity $\sigma$ changes by a factor of two between the NaCl solution at saturation and that at $10 \%$ by mass.  All resistances measured should be inversely proportional to $\sigma$, and so to compare the different solutions we looked at $R_{CR}\sigma$.  $R_{CR}\sigma$ versus $\tau$ is plotted in Fig. 12(c), and does not change, within error, as the concentration is varied.

With changing concentration, the fluid parameters also change slightly.  For a solution of $10$  $\%$ NaCl by mass, $\rho = 1.0707$ g/cm$^3$, $\nu = 1.115$ cSt, and $\gamma = 76.05$ dyne/cm.  Calculating the predicted prefactors for the model, we find that the differences due to the fluid parameters are minimal, and we should not see effects from them outside of our experimental error.

\subsection {Drop snap-off}

\begin{figure}
\centering
\includegraphics[width=80 mm]{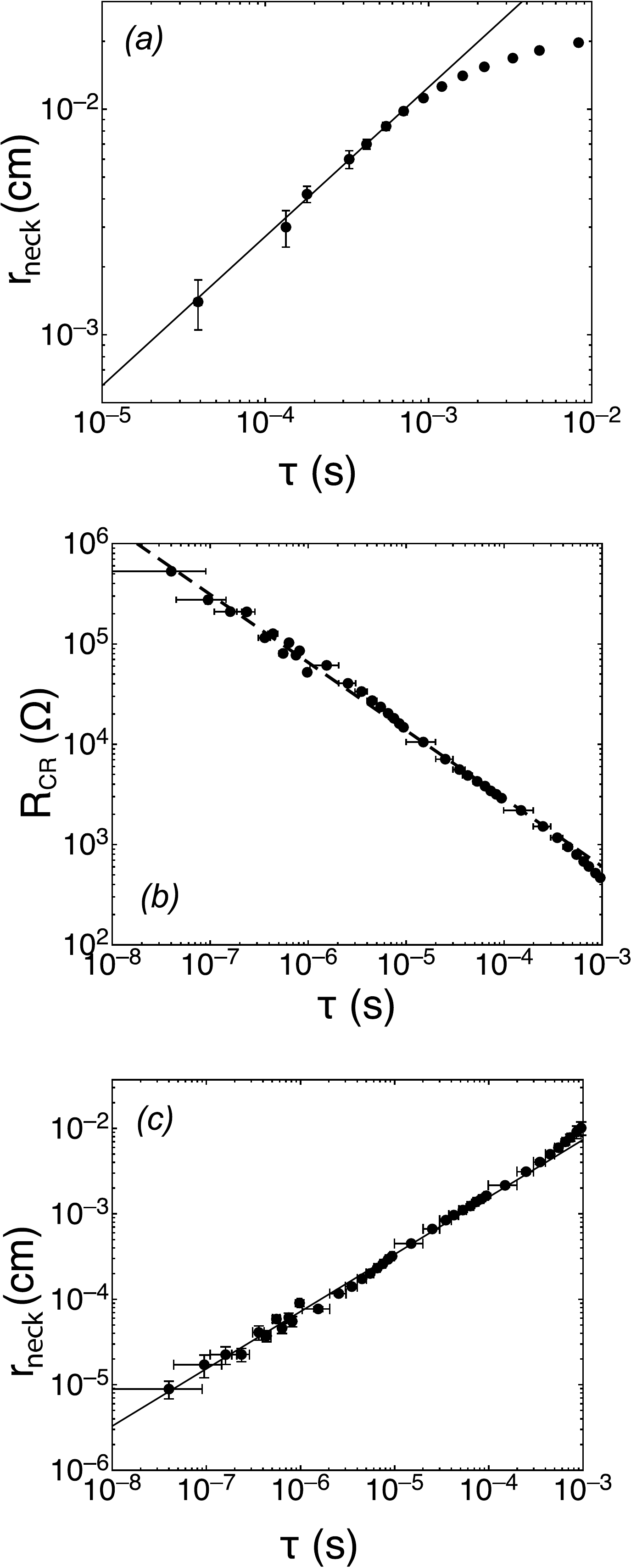}
\caption{Drop snap off.  (a) $r_{neck}$ versus $\tau$.   $r_{neck}$ was measured using high-speed imaging at 144,000 frames per second.  The solid line shows $r = 1.2 \tau^{-0.66}$ (b) $R_{CR}$ versus $\tau$.  The dashed line shows $R_{CR} = 5.7 \tau^{-0.67}$(c) $r_{neck}$ versus $\tau$.  $r_{neck}$ calculated from electrical measurement of $R_{CR}$.  The solid line shows $r_{neck} =0.75 \tau^{-0.67}$.}
\end{figure} 

Finally, we used our method to study drop break-up.  We compare the output of our method against previous work, as well as our own calculations, and find that they are consistent.

It has been previously observed that during drop snap-off, the neck between the two drops forms a self-similar cone with an angle of  approximately $18^o$.  We numerically calculated the resistance of a truncated cone of fixed larger radius as a function of the smaller radius, $r_{neck}$, and found $R_{CR} = 1.18/\sigma r_{neck}$ where $\sigma$ is in units of 1/$\Omega \cdot$ cm.  
 
 We used a Phantom V.7 fast digital camera at 144,000 frames per second to measure $r_{neck}$ as a function of $t - t_0 = \tau$, where $t_0$ is the instant of snap off.  As shown in Fig. 13(a), the best fit to our data with $\tau < 1$ ms yields $r_{neck} = (1.2 \pm 0.6) \tau^{-0.66 \pm 0.06}$, which is in agreement with previous measurements of drop break-up of water in air.  

As shown in Fig. 13(b), $R_{CR}$ versus $\tau$ for a snap-off event yields a best fit of $R_{CR} = (5.7 \pm 0.9) \tau^{-0.67 \pm 0.01}$.  This is shown as the dashed line in the figure.  Combining the electrostatic calculation with the measured dependence of $r_{neck}$ on $\tau$, we predict that $R_{CR} = (5.24 \pm 2.63) \tau^{-0.66 \pm 0.06}$, within error of the data.

This indicates that the scaling law for break up of water in air persists to timescales of 10 ns.  Using the relation above, we calculate $r_{neck}$ from $R_{CR}$.  $r_{neck}$ versus $\tau$ is shown in Fig. 13(c).  The solid line is the best fit to the data, $r_{neck} = (0.75 \pm 0.01) \tau^{-0.67 \pm 0.01}$.

\section{Capacitance Measurements}

Before the bridge is formed, we need to separate the capacitance of the coalescing region, $C_{CR}$ from the capacitance of the total arrangement of conductors.  As stated in the experimental section, in order to achieve this, I measure the capacitance of the cell before any drops are formed on the nozzle tips, $C_{cell}$, and subtract this from the $C_{CR}$ measured in the last $\mu$s before coalescence occurs.

I justify this approximation by simulating the axially symmetric part of the cell, including the nozzles and the tubes filled with NaCl solution.  I find that for a fixed separation of the two drops ($A = 1$ mm) of $d = 1.5$ mm, $C_{CR} = 0.45$ pF from the simulation.  Simulating the cell with no drops formed gives $C_{cell} = 0.44$ pF.  

For a $1.5$ mm separation between the drop tips, the measurements show $C_{CR} = 0.66 \pm 0.02$ pF, and $C_{cell} = 0.62 \pm 0.02$ pF.  When there is no fluid in the cell, we measure a stray capacitance of $0.2$ pF, which accounts for the difference between the measurement and simulation if it can be considered to be in parallel with $C_{CR}$.

The capacitance of the cell outside the ``coalescing region'' is in parallel with $C_{init}$, thus, I estimate that $C_{init} = C_{CR} - C_{cell}$.  The simulation when $d = 1.5$ mm yields $C_{CR} - C_{cell} = 0.01$ pF, which is consistent with our measurement of $C_{CR} - C_{cell} = 0.04 \pm 0.03$ pF.  

In order to find $C_{init}$ just before coalescence, I measure $C_{cell}$ at the nozzle separation used for the coalescence measurements, and subtract this from $C_{CR}$.  For $A = 1$ mm and a nozzle separation of 2 mm, $C_{cell} = 0.89$ pF.  The simulation predicts $C_{cell} = 0.64$ pF in this case.  This measurement is repeated for $A = 2$ mm and $d = 4$ mm, where $C_{cell} = 0.92$ pF, and at $A = 0.75$ mm and $d = 1.5$ mm, where $C_{cell}$ = $0.81$ pF.

\end{document}